\documentclass{elsart3}
\usepackage[german,english]{babel}
\usepackage{epsfig}
\usepackage{picins}
\usepackage[centertags]{amsmath}
\usepackage{amssymb,amscd}
\usepackage{./styles/mcite}
\usepackage{./styles/myxspace}
\def\zero{{\scriptscriptstyle 0}}

\catcode`\@=11 
\@ifundefined{euro}%
{\providecommand{\Euro}{Euro\xspace}
 }%
{\providecommand{\Euro}{\euro\xspace}
 }
\catcode`\@=12 

\def\Z0{{Z^\zero}}
\def\eVdist{\kern-0.06667em}

\def\ev{{\,\text{e}\eVdist\text{V\/}}}

\def\gev{{\,\text{Ge}\eVdist\text{V\/}}}
\def\tev{{\,\text{Te}\eVdist\text{V\/}}}

\def\IP{{\rm I$\kern-0.01667em$P}\xspace}

\def\Ptlj{{\not{\kern-0.55ex P}}_t\ell j}
\def\Ptmiss{{\not{\kern-0.55ex P}}_t}

\def\rnge{{\,\text{--}\,}}

\newcommand{\nucl}[2]{\ensuremath{\stru{2.ex}{0.8ex}^{#2}{\rm #1}}}

\mathchardef\qsm=63
\mathchardef\pls=43
\mathchardef\mns=512
\mathchardef\plm=518
\mathchardef\eql=61
\mathchardef\smallleft=300
\mathchardef\smallright=301
\mathchardef\perslsh=47
\mathchardef\les=316
\mathchardef\gre=318
\mathchardef\leq=532
\mathchardef\grq=533
\chardef\usc=95
\chardef\til=126

\def\sqr#1#2#3{{\vcenter{\hrule height.#3ex\hbox{\vrule width.#2ex height#1ex
    \kern#1ex\vrule width.#3ex}\hrule height.#2ex}}}

\def\angleto{\vrule width.035em height2.1ex depth-.56ex\unskip\kern-.6ex\to}
\def\perchc#1{{\raise.4ex\hbox{$\mkern4mu#1{\it\perslsh}_
             {\mkern-5mu\scriptscriptstyle{{\rm o}\!{\rm o}}}^
             {\mkern-12.8mu\scriptscriptstyle{\rm o}}$}}}

\catcode`\@=11 
\def\parenbar{\mathpalette\p@renb@r}
\def\p@renb@r#1#2{\vbox{%
  \ifx#1\scriptscriptstyle \dimen@.7em\dimen@ii.2em\else
  \ifx#1\scriptstyle \dimen@.8em\dimen@ii.25em\else
  \dimen@1em\dimen@ii.4em\fi\fi \offinterlineskip
  \ialign{\hfill##\hfill\cr
    \vbox{\hrule width\dimen@ii}\cr
    \noalign{\vskip-.3ex}%
    \hbox to\dimen@{$\mathchar300\hfil\mathchar301$}\cr
    \noalign{\vskip-.3ex}%
    $#1#2$\cr}}}
\catcode`\@=12 

\newbox\struttbox
\setbox\struttbox=\hbox{\vrule height1.65ex depth.485ex width0pt}
\def\strutt{\relax\ifmmode\copy\struttbox\else\unhcopy\struttbox\fi}
\def\stru#1#2{\relax\ifmmode\hbox{\vrule height#1 depth#2 width0pt}
\else\vrule height#1 depth#2 width0pt\fi}
\def\uline#1{$\underline{\hbox{#1\strutt}}$}
\def\ronum#1{\uppercase\expandafter{\romannumeral#1}}
\def\ronuml#1{\expandafter{\romannumeral#1}}

\def\cbk{\kern-0.5em}
\newcommand{\pcite}[1]{{\protect\cite{#1}}}

\newcommand{\linebox}[2][3.ex]{\uline{\hbox to #2{\stru{#1}{0.pt}\hfil}}}

\newcounter{seqnum}
\DeclareMathAlphabet{\mathbf}{OT1}{cmr}{bx}{n}

\DeclareMathAlphabet{\mathbfs}{OT1}{lcmss}{bx}{sl}

\newlength\listtextwidth

\catcode`\@=11 
\newlength{\@tabfninsert}
\newlength{\@tabfnwidth}
\newcommand{\tabfootnote}[2]{%
  \setlength{\@tabfninsert}{0.8em}
  \setlength{\@tabfnwidth}{\textwidth}
  \addtolength{\@tabfnwidth}{-\@tabfninsert}
  \addtolength{\@tabfnwidth}{-0.4em}
  \noindent\makebox[\@tabfninsert][r]{\footnotesize$^{#1}$\hfil}\hfill%
  \parbox[t]{\@tabfnwidth}{\footnotesize #2\hfill}}
\catcode`\@=12 
\newcommand{\boldarrayrulewidth}{1pt}

\catcode`\@=11 
\let\tab@penalty\relax
\newcount\tab@state
\def\bcline#1{%
  \noalign{\kern-.5\arrayrulewidth\tab@penalty}%
  \omit%
  \global\tab@state\@ne%
  \ranges\bcline@i{#1}%
  \cr%
  \noalign{\kern-.5\arrayrulewidth\tab@penalty}%
}
\def\bcline@i#1#2{%
  \ifnum#1<\tab@state\relax%
    \tab@@cr%
    \noalign{\kern-\arrayrulewidth\tab@penalty}%
    \omit%
    \global\tab@state\@ne%
  \fi%
  \@whilenum\tab@state<#1\do{%
    \hfil\tab@@tab@omit%
    \global\advance\tab@state\@ne%
  }%
  \ifnum\tab@state>\@ne%
    \kern-\arrayrulewidth%
  \fi%
  \@whilenum\tab@state<#2\do{%
    \tab@@span@omit%
    \global\advance\tab@state\@ne%
  }%
  \leaders\hrule\@height\boldarrayrulewidth\hfill%
}
\def\ranges#1#2{%
  \gdef\ranges@temp{#1}%
  \begingroup%
  \ranges@i#2 \q@delim%
}
\def\ranges@i{%
  \@ifnextchar\q@delim\ranges@done{\afterassignment\ranges@ii\count@}%
}
\def\ranges@ii{%
  \@ifnextchar-\ranges@iii{\ranges@do\count@\count@\ranges@v}%
}
\def\ranges@iii-{\afterassignment\ranges@iv\@tempcnta}
\def\ranges@iv{\ranges@do\count@\@tempcnta\ranges@v}
\def\ranges@v{%
  \@ifnextchar,%
    \ranges@vi%
    {%
      \@ifnextchar\q@delim%
        \ranges@done%
        {\tab@err@range\ranges@vi,}%
    }%
}
\def\ranges@vi,{\afterassignment\ranges@ii\count@}
\def\ranges@do#1#2{%
  \ifnum#1>#2\else%
    \expandafter\endgroup%
    \expandafter\ranges@temp%
    \expandafter{%
    \the\expandafter#1%
    \expandafter}%
    \expandafter{%
    \the#2%
    }%
    \begingroup%
  \fi%
}
\def\ranges@done\q@delim{\endgroup}
\def\ifinrange#1#2{%
  \@tempswafalse%
  \count@#1%
  \ranges\ifinrange@i{#2}%
  \if@tempswa%
    \expandafter\@firstoftwo%
  \else%
    \expandafter\@secondoftwo%
  \fi%
}
\def\ifinrange@i#1#2{%
  \ifnum\count@<#1 \else\ifnum\count@>#2 \else\@tempswatrue\fi\fi%
}
\def\tab@@cr{\cr}
\def\tab@@tab@omit{&\omit}
\def\tab@@span@omit{\span\omit}
\def\tab@checkrule#1{%
  \count@#1\relax%
  \expandafter\ifinrange%
  \expandafter\count@%
  \expandafter{\tab@xcols}%
    {\tab@checkrule@i}%
    {}%
}
\def\bhline{\noalign{\ifnum0=`}\fi\hrule \@height  
\boldarrayrulewidth \futurelet \@tempa\@xhline}
\def\@xhline{\ifx\@tempa\hline\vskip \doublerulesep\fi
      \ifnum0=`{\fi}}
\catcode`\@=12 

\catcode`\@=11 
\newcounter{pict@width}
\newcounter{pict@height}
\newlength{\pict@scale}
\newcommand{\psfigadd}[4]{%
\setcounter{pict@width}{1*\ratio{#2+\pict@scale/2}{\pict@scale}}
\setcounter{pict@height}{1*\ratio{#3+\pict@scale/2}{\pict@scale}}
\setlength{\unitlength}{\pict@scale}
\hbox to #2{\hspace{-\fill}\begin{picture}(\thepict@width,\thepict@height)
\put(0,0){\psfig{figure=#1,width=#2,height=#3,clip=}}
\SetScale{0.283466457}
\SetWidth{1.763889}
{#4}
\end{picture}}
}
\newcounter{pict@widthfst}
\newcounter{pict@widthscd}
\newcounter{pict@widthtot}
\newcommand{\psfigaddtwo}[7]{%
\setcounter{pict@widthfst}{1*\ratio{#2+\pict@scale/2}{\pict@scale}}
\setcounter{pict@widthscd}{1*\ratio{#2+#4+\pict@scale/2}{\pict@scale}}
\setcounter{pict@widthtot}{1*\ratio{#2+#4+#6+\pict@scale/2}{\pict@scale}}
\setcounter{pict@height}{1*\ratio{#3+\pict@scale/2}{\pict@scale}}
\setlength{\unitlength}{\pict@scale}
\hbox{\hspace{-\fill}\begin{picture}(\thepict@widthtot,\thepict@height)
\put(0,0){\psfig{figure=#1,width=#2,height=#3,clip=}}
\put(\thepict@widthscd,0){\psfig{figure=#5,width=#6,height=#3,clip=}}
\SetScale{0.283466457}
\SetWidth{1.763889}
{#7}
\end{picture}}
}
\newcommand{\psfigror}[4]{%
\setcounter{pict@width}{1*\ratio{#2+\pict@scale/2}{\pict@scale}}
\setcounter{pict@height}{1*\ratio{#3+\pict@scale/2}{\pict@scale}}
\setlength{\unitlength}{\pict@scale}
\hbox{\begin{picture}(\thepict@width,\thepict@height)
\put(0,\thepict@height){\psfig{figure=#1,width=#3,height=#2,clip=,angle=270}}
\SetScale{0.283466457}
\SetWidth{1.763889}
{#4}
\end{picture}}
}
\newcommand{\psfigrol}[4]{%
\setcounter{pict@width}{1*\ratio{#2+\pict@scale/2}{\pict@scale}}
\setcounter{pict@height}{1*\ratio{#3+\pict@scale/2}{\pict@scale}}
\setlength{\unitlength}{\pict@scale}
\hbox{\begin{picture}(\thepict@width,\thepict@height)
\put(0,0){\psfig{figure=#1,width=#3,height=#2,clip=,angle=90}}
\SetScale{0.283466457}
\SetWidth{1.763889}
{#4}
\end{picture}}
}
\catcode`\@=12 

\usepackage{ulem}
\catcode`\@=11\def\@date{14th April 2006}\catcode`\@=12
\begin{document}
\begin{frontmatter}



\title{
\vspace*{-2.cm}
\rightline{\normalsize\bf FAU-PI1-06-01}
\vspace*{1.5cm}
KM3NeT: Towards a km$^3$ Mediterranean Neutrino Telescope}


\author{U.F.~Katz}
\ead{katz@physik.uni-erlangen.de}
\address{University of Erlangen-Nuremberg, Physics Institute}

\begin{abstract}
The observation of high-energy extraterrestrial neutrinos is one of the most
promising future options to increase our knowledge on non-thermal processes in
the universe. Neutrinos are e.g.\ unavoidably produced in environments where
high-energy hadrons collide; in particular this almost certainly must be true in
the astrophysical accelerators of cosmic rays, which thus could be identified
unambiguously by sky observations in ``neutrino light''. To establish neutrino
astronomy beyond the detection of single events, neutrino telescopes of km$^3$
scale are needed. In order to obtain full sky coverage, a corresponding detector
in the Mediterranean Sea is required to complement the IceCube experiment
currently under construction at the South Pole. The groups pursuing the current
neutrino telescope projects in the Mediterranean Sea, ANTARES, NEMO and NESTOR,
have joined to prepare this future installation in a 3-year, EU-funded Design
Study named KM3NeT (in the following, this name will also denote the future
detector). This report will highlight some of the physics issues to be addressed
with KM3NeT and will outline the path towards its realisation, with a focus on the
upcoming Design Study.
\end{abstract}

\end{frontmatter}
\section{Physics with KM3NeT}
\label{sec-phy}

The energy range accessible to neutrino telescopes is intrinsically limited by
the detection method to some $10\gev$ at its lower end, while at energies beyond
roughly $10^{17}\ev$ the neutrino flux is expected to fade below detection
thresholds even for future km$^3$-scale detectors. The lower-energy region is
dominated by the flux of {\it atmospheric neutrinos} (cf.\ Fig.~\ref{fig-difl}),
produced in reactions of cosmic rays with the atmosphere. There are three
approaches to identify cosmic muon signals on top of this background:
\begin{enumerate}
\item
Neutrinos from specific astrophysical objects ({\it point sources})
produce excess signals associated to particular celestial coordinates.
\item
Neutrinos not associated to specific point sources ({\it diffuse flux}) are
expected to have a much harder energy spectrum than the atmospheric neutrinos
and to dominate the neutrino flux above $10^{14}\rnge10^{15}\ev$.
\item
Exploitation of coincidences in time and/or direction of neutrino events with
observations by telescopes (e.g.\ in the radio, visible, X-ray or gamma regimes)
and possibly also by cosmic ray detectors ({\it multimessenger method}).
\end{enumerate}

The various astro- and particle physics questions to be addressed with the
resulting data have been summarised e.g.\ in \cite{astro-ph-0503122} and
references therein. Here, we will focus on a few central topics, including a
recent development:

\subsection{Neutrinos from galactic shell-type supernova remnants}
\label{sec-phy-snr}

\begin{figure}[b]
  \begin{center}
  \epsfig{file=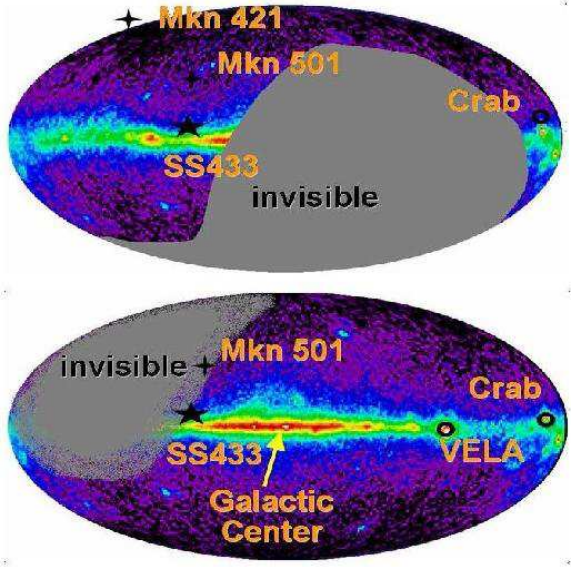,width=\columnwidth}
  \end{center}
  \caption{Field of view of a neutrino telescope at the South Pole (top) and in
           the Mediterranean (bottom), given in galactic coordinates. A
           $2\pi$-downward sensitivity is assumed; the gray regions are then
           invisible. Indicated are the positions of some candidate neutrino
           sources.}
  \label{fig-fov}
\end{figure}

The shock waves developing when supernova ejecta hit the interstellar medium
are prime candidates for hadron acceleration through the Fermi mechanism. Recent
observations of gamma rays up to energies of about $40\tev$ from two shell-type
supernova remnants in the Galactic plane (RX\,J1713.7-3946 and RX\,J0852.0-4622)
\cite{astro-ph-0511678,aa:437:l7} with the H.E.S.S.\ \v Cerenkov telescope
support this hypothesis and disfavour explanations of the gamma flux by purely
electromagnetic processes. The detection of neutrinos from these sources would,
for the first time, identify unambiguously specific cosmic accelerators. Note
that this is only possible with Northern-hemisphere neutrino telescopes which,
in contrast to the South Pole detectors, cover the relevant part of the Galactic
plane in their field of view (see Fig.~\ref{fig-fov}).

\begin{figure}[hbt]
  \epsfig{file=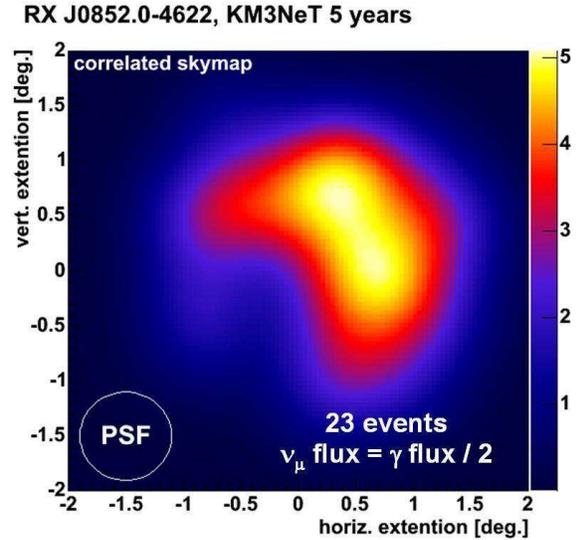,width=\columnwidth,%
          bblly=24,bbllx=19,bbury=552,bburx=570,clip=}
  \caption{\protect\raggedright
           A skymap of the simulated neutrino signal from RX\,J0852.0-4622 as
           seen by a km$^3$-scale neutrino telescope in the Mediterranean Sea
           after 5 years of data taking. In the simulation, a power-law gamma
           spectrum without energy cut-off and the relation
           $\phi_{\nu_\mu}/\phi_\gamma=1/2$ have been assumed. The background of
           atmospheric neutrinos, not included in the plot, can be efficiently
           reduced by adjusting the lower energy cut without affecting
           significantly the signal. The circle in the lower left corner
           indicates the average angular resolution (point spread function).}
  \label{fig-km3hess}
\end{figure}

The expected event rates can be estimated using the rough assumptions that the
gamma flux follows a power-law spectrum without high-energy cut-off and the muon
neutrino and gamma fluxes are in relation $\phi_{\nu_\mu}/\phi_\gamma=1/2$,
taking into account the relative production probabilities of charged and neutral
pions, their decay chains and neutrino oscillations. Preliminary calculations
indicate that the first-generation Mediterranean neutrino telescopes may have a
chance to observe a few events, whereas a significantly larger signal is
expected in a future cubic-kilometre set-up; a tentative estimate of the
neutrino sky map of RX\,J0852.0-4622 after 5~years of data taking with KM3NeT is
shown in Fig.~\ref{fig-km3hess}.

\subsection{The diffuse neutrino flux}
\label{sec-phy-dif}

\begin{figure}[htb]
  \epsfig{file=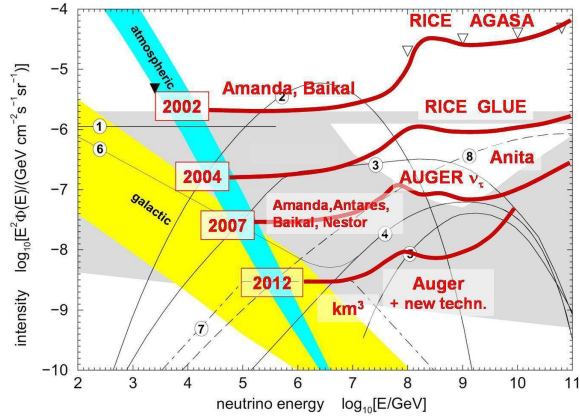,width=\columnwidth,%
          bblly=24,bbllx=19,bbury=428,bburx=572,clip=}
  \caption{Experimental sensitivity to the diffuse neutrino flux for various
           current and future experiments (red lines), compared to different
           models for contributions to the diffuse flux (numbered lines). See
           \pcite{arevns:g29:843} for detailed explanations. The flux of
           atmospheric neutrinos is indicated as blue band. Plot provided by
           courtesy of C.~Spiering.
           \label{fig-difl}}
\end{figure}

The sensitivity of current and future experiments is compared to various
predictions of diffuse neutrino fluxes in Fig.~\ref{fig-difl} (following
\cite{jp:g29:843,arevns:g29:843}). Whereas some of the models are already now
severely constrained by the data, others require km$^3$-size neutrino telescopes
for experimental assessment and potential discoveries. The measurement of the
diffuse neutrino flux would allow for important clues on the properties of the
sources, on their cosmic distribution, and on more exotic scenarios such as
neutrinos from decays of topological defects or superheavy particles ({\it
top-down scenarios}).

\subsection{Search for dark matter annihilation}
\label{sec-phy-dms}
The major part of the matter content of the universe is nowadays thought to be
formed by {\it dark matter}, i.e.\ non-baryonic, weakly interacting massive
particles (WIMPs); an attractive WIMP candidate is the lightest supersymmetric
particle, the neutralino. Complementary to direct searches for WIMPs, indirect
WIMP observations could be possible by measuring neutrinos produced in WIMP
annihilation reactions in regions with enhanced WIMP density. Such accumulations
may in particular occur due to gravitational trapping, e.g.\ in the Sun or the
Galactic Centre.

The WIMP signal would be an enhanced neutrino flux from these directions, with a
characteristic upper cut-off in the energy spectrum below the WIMP mass,
$M_\text{WIMP}$. Although there is no generic upper constraint on
$M_\text{WIMP}$, supersymmetric theories prefer values below $1\tev$. 
Substantial detection efficiency down to order $100\gev$ is therefore essential
for indirect WIMP searches through neutrinos. The expected sensitivity depends
strongly on assumptions on the WIMP density profile, on $M_\text{WIMP}$ and on
the energy spectrum of neutrinos from WIMP annihilations. At least for some
supersymmetric scenarios this sensitivity is compatible or even better than for
direct searches \cite{astro-ph-0503122}.

\section{Requirements for KM3NeT}
\label{sec-req}

In 2002 the {\it High Energy Neutrino Astronomy Panel (HENAP)} of the
PaNAGIC\footnote{Particle and Nuclear Astrophysics and Gravitation
International Committee} Committee of {IUPAP}\footnote{International Union of
Pure and Applied Physics} has concluded that ``a km$^3$-scale detector in the
Northern hemisphere should be built to complement the IceCube detector being
constructed at the South Pole'' \cite{misc:henap:2002}; one major argument in
favour of this effort is the coverage of the Southern sky including the central
part of the Galactic plane (cf.\ Fig.~\ref{fig-fov} and Sect.~\ref{sec-phy}).

The major challenges in constructing such a deep-sea neutrino telescope are the
high pressure of several $100\,$bar; the uncontrollable environment with
currents, sedimentation and background light from $\nucl{K}{40}$ decays and
bioluminescent organisms; the chemically aggressive environment reducing the
selection of suited materials basically to titanium, glass and certain
plastics. A further aspect of this difficult environment is that the deployment
and maintenance operations, employing surface vessels and manned or
remotely-operated deep-see submersibles, are expensive and weather-dependent,
thus maximising the need for high operation stability.

These issues have been sucessfully addressed by the Mediterranean neutrino
telescope projects ANTARES \cite{astro-ph-9907432,misc:ant:tdr} (under
construction near Toulon), NEMO \cite{npps:143:359,misc:nemo-home} (R\&D for a
cubic kilometre detector) and NESTOR \cite{ncim:24c:771,proc:icrc:2003:1377}
(under construction near the west coast of the Pelopones). Suitable technical
solutions have been identified and tested in the context of these
first-generation projects, and the feasibility of large deep-sea detectors has
been proven. Nevertheless, further R\&D work is necessary for KM3NeT, a.o.\ in
order to
\begin{itemize}
\item
improve cost effectiveness;
\item
optimize the geometrical layout of a km$^3$-scale detector and the
corresponding mechanical structures;
\item
find the optimal combination of photosensor, readout and data acquisition
options;
\item
define the deep-sea infrastructure and the deployment and maintenance
procedures;
\item
provide suitable interfaces addressing the needs of the deep-sea science
communities.
\end{itemize}

The instrumented water volume of KM3NeT is thought to be about one km$^3$,
which is the ``canonical size'' to cover the physics objectives described
above.  However, it is anticipted that the results of this detector may imply
the need for increasing the sensitivity further, thus requiring an extension in
volume. It is therefore considered an intrisic requirement for the KM3NeT
design to be extendable.

An important constraint is imposed by the timeline of the IceCube
neutrino telescope \cite{misc:icube:pdd} currently being installed in
the Antarctic ice. IceCube will be completed by 2011, and it is a
major objective for KM3NeT to be operational in time to take data
concurrently with IceCube.

\section{The KM3NeT Design Study}
\label{sec-des}

The 6th Framework Programme of the European Union offers support for the
development of future research infrastructures by funding {\it Design Studies}
as {\it Specific Support Actions}. A corresponding application has been
submitted in March 2004 and was approved in September 2005. The Design Study
project has started in February 2006 and will run for 3 years. It is funded
with 9\,million \Euro by the EU and has an overall volume of about 20\,million
\Euro, which will mainly be used for personnel and costs for prototyping,
deployment tools and tests, etc.

\subsection{The KM3NeT Consortium}
\label{sec-des-con}

The KM3NeT consortium is composed of 29 particle/astroparticle physics and 7 sea
science/ technology institutes from Cyprus, France, Germany, Greece, Italy, the
Netherlands, Spain and the United Kingdom. The Design Study comprises, amongst
others, the groups involved in ANTARES, NEMO and NESTOR, and is coordinated by
the University of Erlangen, Germany.

\subsection{Objectives of the Design Study}
\label{sec-des-obj}

The major objective of the KM3NeT Design Study is to work out the technical
foundation for the construction of the neutrino telescope, to be documented in
a {\it Technical Design Report (TDR)}. The TDR and and an intermediate
{\it Coceptual Design Report (CDR)} will be the main deliverables of the 
Design Study. Their preparation will require 
\begin{itemize}
\item
a critical review of current technical solutions;
\item
development and thorough tests of new solutions;
\item
the assessment of quality control and assurance;
\item
the exploration of links to industry, in particular in the fields of
photodetection, information technology and deep-sea technology;
\item
careful studies of the interrelation between the different aspects and
the optimisation of the solutions found.
\end{itemize}

The goal is to design a neutrino telescope with sensitivity down to neutrino
energies $E_\nu$ of a few $100\gev$. The low level of light scattering in
deep-sea water is to be exploited to reach a pointing resolution limited by the
average angle between incoming neutrino and secondary muon up to
$E_\nu\sim10\tev$ and better than $0.1^\circ$ above this energy.  The vision of
the proponents is that KM3NeT will be a pan-European research infrastructure,
giving open access to the neutrino telescope data, allowing to assign
``observation time'' to external users by adapting the online filter algorithms
to be particularly sensitive in predefined celestial directions, and providing
access to long-term deep-sea measurements to the marine sciences communities.

Three possible sites for the KM3NeT infrastructure have been identified: the
ANTARES site near Toulon, the NEMO site near Capo Passero in Sicily and the
NESTOR site near Pylos on the Pelopones. The existing studies on the
characteristics of these sites (water transparency, currents, sedimentatiion,
bioluminescence, etc.) will be consolidated during the Design Study, and the
site parameters (depth, distance to shore, etc.) will be taken into account in
the optimisation of the physics sensitivity, where the figure of merit will be
{\it physics output per \Euro}. The results of these studies will provide the
scientific input to the site decision, which will be a political process to be
initiated towards the end of the Design Study.

In parallel with preparing the site decision process, also funding and
governance models and options will be investigated, so that the corresponding
decisions can be taken in due time.

\subsection{Some Key Questions to be Addressed}
\label{sec-des-que}

The list of requirements given in Sect.~\ref{sec-req} directly translates into
major activities during the Design Study, of which a few key examples are to be
discussed in the following.

\smallskip
\uline{Detector architecture:} 
Although all first-gen\-er\-a\-tion projects use large-diameter photomultipliers
(PMs), they pursue different approaches for the arrangement and mechanical
support of the PMs. A choice has to be made between flexible stings (like in
ANTARES), towers formed by rigid structures (NESTOR), towers formed by rigid
arms connected to each other by ropes forming a tetrahedral structure (NEMO), a
combination of these, or yet other solutions. This question is closely related
to a variety of aspects, like
\begin{itemize}
\item
the physics sensitivity, which strongly depends on the geometrical arrengement
of the photosensors (see also \cite{proc:vlvnt2:2005:skuch});
\item
the dee-sea infrastructure (cables, power distribution, data transport);
\item
the deployment procedures;
\item
the calibration methods;
\item
readout and data acquisition.
\end{itemize}

This example demonstrates how strongly the different technical issues are
interrelated. It will be an essential part of the Design Study to assess these
interdependencies.

\begin{figure}[htb]
  \begin{center}
  \epsfig{file=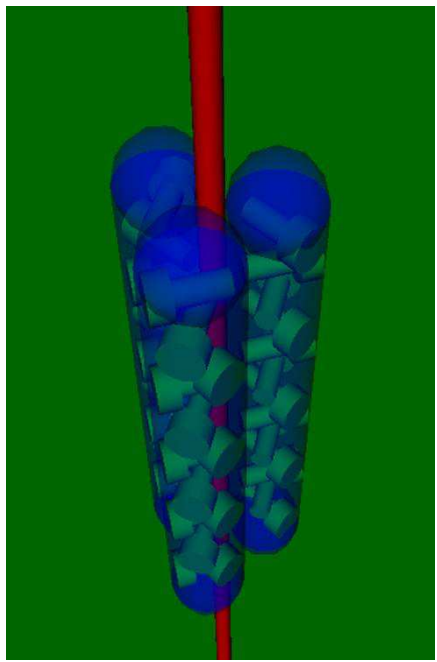,width=0.75\columnwidth}
  \end{center}
  \caption{Sketch of a possible arrangement of several photomultiplier tubes
           in three glass cylinders.
           \label{fig-omc}}
\end{figure}

\smallskip
\uline{Photodetection:} 
The KM3NeT timeline (see Sect.~\ref{sec-pat}) is too tight to embark on the
development of alternative photodetection methods, such as HPDs or silicon
PMs. Nevertheless, there are different options (again having specific
implications for various aspects) that have to be assessed. It has e.g.\ been
suggested to use smaller PMs arranged in cylindrical glass vessels (see
Fig.~\ref{fig-omc} for an illustration). Careful studies are required to
investigate the physics sensitivity of such arrangements
\cite{proc:vlvnt2:2005:skuch}, to optimise the readout, to assess the
implications for the overall detector cost, etc.

\smallskip
\uline{Deployment and deep-sea infrastructure:}
For the deployment of the detector modules, different approaches have been
developed by the first-generation projects (e.g.\ connections performed by
remotely-operated submersibles or at surface, deploymnt from ships or dedicated
platforms, etc.) which need to be adapted to the needs of the KM3NeT
infrastructure. Further issues to be addressed include the architecture of the
deep-sea cable net, including the choice of the components and the installation
and maintenance procedures.

\section{The Path to KM3NeT Construction}
\label{sec-pat}

The KM3NeT Design Study will last until January 2009. Thereafter, a phase of
funding negotiations and construction preparation has to be foreseen, lasting
1--2 years. This phase might be supported within the 7th Framework Programme of
the European Union; an important step into ths direction is the inclusion of
the KM3NeT project into the {\it List of Opportunities}
\cite{misc:esfri-loo:2005} of the {\it European Strategy Forum for Research
Infrastructures (ESFRI)}. If the decision to realise the KM3NeT infrastructure
is taken in this preparation phase, installation could start as early as 2010
and be concluded in 2012. First data would thus become available in 2011,
concurrently with data from the IceCube telescope.

\section{Conclusions}
\label{sec-con}

Neutrino astronomy is an emerging field in astroparticle physics offering
exciting prospects for gaining new insights into the high-energy, non-thermal
processes in our universe. The current neutrino telescope projects in the
Mediterranean Sea are approaching installation and promise exciting first data. 
They have reached a level of technical maturity allowing for the preparation of
the next-generation cubic-kilometre detector to complement the IceCube telescope
currently being installed at the South Pole. The interest in this project has
been further enhanced by the recent H.E.S.S.\ observations of high-energy gamma
rays from shell-type supernova remnants in the Galactic plane, indicating that
these objects could well be intense neutrino sources, which would, however, be
invisible to IceCube.

The technical design of the future Mediterranean km$^3$ neutrino telescope will
be worked out in the 3-year EU-funded KM3NeT Design Study starting in February
2006. The construction of the KM3NeT neutrino telescope during the first years
of the next decade thus appears to be possible.

\vspace*{3.mm}

\noindent
{\bf Acknowledgement:} The author wishes to thank the organisers of the VLVnT2
Workshop in Catania for their overwhelming hospitality and a very intense,
productive and perfectly organised workshop.

{
\bibliographystyle{./Catania}
{\raggedright\fontsize{10.pt}{11.pt}\selectfont
\bibliography{./Catania.bib}}
}

\end{document}